\documentclass{ws-procs975x65}
\newcommand\g{\ensuremath{\mathfrak{g}}}
\newcommand\del{\ensuremath{\nabla}\!}

\begin{document}
\title{ACCELERATED EXPANSION BY NON-MINIMALLY COUPLED SCALAR FIELDS}
\author{ROGER BIELI}
\address{Max Planck Institute for Gravitational Physics\\
Am M\"uhlenberg 1, 14476 Golm, Germany}

\wstoc{ACCELERATED EXPANSION BY NON-MINIMALLY COUPLED SCALAR FIELDS}{BIELI,
ROGER}

\begin{abstract}
In a class of spatially homogeneous cosmologies including those of Bianchi
type I--VIII mathematical results are presented which show that a scalar
field non-minimally coupled to the scalar curvature of spacetime can
dynamically yield a positive cosmological constant without the potential
being required to include one. More precisely, it is shown that in an
exponential potential any positive coupling constant leads eventually
to late-time de Sitter expansion and isotropization corresponding to a
positive cosmological constant and that this behaviour is independent
of the steepness of the potential. This is in marked contrast to the
minimally coupled case where power-law inflation occurs at most, provided
the potential is sufficiently shallow. 
\end{abstract}

\keywords{Dark energy; quintessence; non-minimal coupling.}

\bodymatter

\section{Introduction}\label{intro}

An ongoing major effort in theoretical cosmology is to satisfactorily
explain the observed late-time acceleration of the universe. For an
explanation to be considered satisfactory the underlying model together
with its parameters is commonly required to come with good motivation from
particle physics theories. This is because, up to now, all observational
data are fully consistent with the $\Lambda$CDM concordance model having a
positive cosmological constant driving the expansion. Every other model that
is capable of reproducing this simple behaviour, \emph{e.g.} a scalar field
in a potential with a positive lower bound \cite{Ren04}, can presently not be
ruled out experimentally and cosmologists are therefore endeavoring to find
models that naturally originate from fundamental physics. However, it seems
quite intricate to construct a viable particle physics theory that yields
a positive cosmological constant --- or, more generally, a non-vanishing
lower bound on the potential of some effective field --- comparable to the
observed value. It is the aim of this proceedings contribution to give an
example of a cosmological model that is well-motivated from a quantum gravity
point of view and includes a scalar field that drives de Sitter expansion at
late times. The potential itself does not contain a cosmological constant,
rather, one is established dynamically by a non-minimal coupling of the
field to the spacetime scalar curvature. This mechanism has previously
been considered heuristically by Tsujikawa \cite{Tsu00} in an inflationary
context for the purpose of sustaining inflation in steep potentials. Here,
mathematical results from Ref.~\refcite{Bie06} are presented that generalize
these findings in particular to spatially homogeneous models of Bianchi
type I--VIII and apply them to obtain late-time exponential acceleration.

\section{The model}\label{model}

A direct coupling of a real scalar field $\phi$ to the spacetime scalar
curvature $R$ can be realized by an additional term $-\xi R \phi^2/2$ in
the action
\[ \mathcal{S} = \int \left[ \frac12 (1-\xi \phi^2)\,{\star R} - \frac12
d\phi \wedge \star d\phi - \star V(\phi) \right] + \mathcal{S}_{\rm m} \]
where $\star$ is the Hodge dual, $V$ the scalar field potential and
$\mathcal{S}_{\rm m}$ the action for all other forms of matter which are
supposed to satisfy the dominant and strong energy conditions. Such a
term arises regularly in quantum field theory in curved spacetimes on
renormalizability grounds\cite{Cal70}.  Because the coupling constant
$\xi$ is dimensionless it persists in the classical limit and it is thus
sensible to include the coupling term already in the effective action. If
the dimension of spacetime is $n+1$ with $n\geq3$, the prominent case of
conformal coupling corresponds to $\xi=(n-1)/4n$.

What remains to be specified is the potential $V$ for the scalar field. It
will be taken exponential, $V(\phi) = \lambda \exp( -\kappa \phi)>0$,
with $\kappa >0$ being referred to as the steepness of the potential. The
reason for taking simple exponentials is that, on one hand, they are easily
motivated as originating from Kaluza--Klein type dimensional reductions
or conformal rescalings and on the other hand, the late-time asymptotics
of the respective models is sufficiently well understood in the minimally
coupled case due to the work of Kitada and Maeda \cite{Kit93}. Specifically,
it is known that all initially expanding solutions of Bianchi type I--VIII
approach a power-law inflationary solution if $0<\kappa<\sqrt2$ and that
the field $\phi$ diverges to infinity.

\section{Late-time asymptotics}\label{asymp}

Suppose that there is an $n$-dimensional Lie group $G$ acting simply
transitively on the surfaces of spatial homogeneity such that the spacetime
$M$ can be written as $M=G \times I$ where $I=\left[t_0,\infty\right[$
shall be an interval of comoving time not bounded from above. Assume that
$G$ allows only for left invariant Riemannian metrics of non-positive scalar
curvature. For $n=3$ this restricts to groups of Bianchi type I--VIII. Let
a metric $g$, an energy-momentum tensor $T$ and a scalar field $\phi$
be given as families $\gamma \in C^2(I,T^0_2\g)$, $\rho \in C^1(I)$,
$j \in C^1(I,T^0_1\g)$, $S \in C^1(I,T^0_2\g)$ and $\phi \in C^2(I)$
of tensors on the Lie algebra $\g$ of $G$ by
\[g=\pi^\ast \gamma - dt\otimes dt,\quad T=\pi^\ast S - \pi^\ast j \otimes dt
- dt\otimes \pi^\ast j + \rho\, dt\otimes dt \]
such that the Einstein--scalar field--matter equations according to the
action above
\begin{align*}
 \begin{split}
  (1-\xi \phi^2)\left( Rc - \frac12 R\, g \right)
  & = (1-2\xi) \del\phi \otimes \del\phi + \left( 2\xi - \frac12 \right)
   |\del\phi|^2_g  g \\
  & \quad - 2\xi \phi \del^{\,2}\phi + 2\xi (\phi\, \Box\phi) g - V(\phi) g + T
 \end{split} \\
 \Box\phi - \xi R\phi - V'(\phi) & = 0 \\
 {\rm div}\, T & = 0
\end{align*}
are satisfied. $Rc$ is the Ricci tensor, $R$ the scalar curvature,
$\del$ the covariant derivative, $\Box$ the covariant wave operator and
$|\cdot|^2_g$ the induced fibre metric, all with respect to $g$. $\pi$ and
$t$ are the canonical projections from M onto its factors and $\pi^\ast$
is the pullback of families of covariant tensors on $\g$ back to $M$. The
Hubble parameter $H$ is given by $H={\rm tr\,}\dot\gamma / 2n$.

If for $\xi>0$, initially $1-\xi\phi^2(t_0)>0$ and $H(t_0)> -[ \log
(1-\xi\phi^2) ] \dot\ (t_0) / (n-1)$ hold, at late times the field $\phi$
approaches
\[ \phi_\infty = \sqrt{\frac1\xi + \frac1{\kappa^2} \left( \frac{n+1}{n-1}
\right)^2 } - \frac1\kappa \frac{n+1}{n-1} \] 
with its first two derivatives vanishing, the Hubble parameter $H$
monotonically decreases towards
\[ H_\infty := \sqrt{\frac2{n(n-1)} \frac{V(\phi_\infty)}{1-\xi
\phi_\infty^2}} \]
and the spatial curvature, the shear as well as matter density $\rho$,
flux $|j|_\gamma$ and pressure ${\rm tr\,} S$ decay exponentially. In
fact, the deceleration parameter $q$ converges to $-1$ and exponential
acceleration and isotropization occurs asymptotically mimicking a positive
cosmological constant of magnitude
\[ \Lambda_{\rm dyn} = \frac{\lambda \exp(-\kappa \phi_\infty)}{1-\xi
\phi_\infty^2}.\]
Note that this is true irrespective of the steepness $\kappa$ of the
potential and for any arbitrarily small positive coupling constant
$\xi$, in whose limit $\xi \to 0$ the expansion $\Lambda_{\rm
dyn}=(1/\sqrt\xi)\exp(-\kappa/\sqrt\xi) (C+O(\sqrt\xi))$ holds for a
constant $C>0$.

The results stated above can be obtained by applying a cosmic no-hair
theorem\cite{Bie06} in a conformally related frame, in which the field
is minimally coupled to gravity but directly coupled to the ordinary
matter instead. It might be interesting that scalar fields non-minimally
coupled to matter were, apart from being conformally related to certain
scalar--tensor theories of gravity, studied \emph{per se} as they can
yield scaling dark energy \cite{Ame99} or hide a putative fifth force from
laboratory experiments \cite{Kho04}. Another consequence of the convergence
of the field $\phi$ to a finite value $\phi_\infty$ worth noting is that,
given the hypotheses above, the quantity $1-\xi \phi^2$ stays bounded
away from zero for all times, a topic relevant for the boundedness of the
effective gravitational constant \cite{Sta81}.

\enlargethispage{\baselineskip}

\end{document}